\documentclass[12pt]{article}

\pdfoutput=1 
\usepackage{hyperref}

\usepackage{caption}
\usepackage{amsmath}
\usepackage{graphicx}
\usepackage{tikz}
\usepackage{epsfig}
\usepackage{dcolumn}
\usepackage{bm}
\usepackage{slashed}
\usepackage{placeins}
\usepackage{float}
\usepackage{amsfonts}
\usepackage{geometry}
\usepackage{xcolor}
\usepackage{cite}
\usepackage{amssymb}%
\usetikzlibrary{automata, positioning, arrows}

 \usepackage{titlesec}
\usepackage{MnSymbol}
\usepackage{wasysym}
\providecommand{\U}[1]{\protect\rule{.1in}{.1in}}

\def\be{\begin{equation}}
\def\ee{\end{equation}}
\def\bea{\begin{eqnarray}}
\def\eea{\end{eqnarray}}
\def\ba{\begin{array}}
\def\ea{\end{array}}
\def\bd{\begin{displaymath}}
\def\ed{\end{displaymath}}

\def\L{\Lambda}

\numberwithin{equation}{section}

\begin{document}

\begin{titlepage}
\hfill LCTP-19-17

\vskip  1cm
\begin{center}
{\Large \bf Microstate Counting via Bethe Ans{\"a}tze }\\

\vskip .7cm

{\Large  \bf in  the 4d ${\cal N}=1$ Superconformal Index}\\
\vskip.7cm

\end{center}

\vskip .7 cm

\vskip 1 cm
\begin{center}
{ \large Alfredo Gonz\'{a}lez Lezcano${}^a$  and  Leopoldo A. Pando Zayas${}^b$}

\vspace{1cm}
{\it ${}^a$ SISSA International School for Advanced Studies}\\
{\it Via Bonomea 265, 34136 Trieste }\\
{\it and }\\
{\it   INFN, sezione di Trieste}\\
\vspace{.4cm}

{\it ${}^{a}$  Departamento de F\'{i}sica,  Universidad de Pinar del R\'{i}o}\\
{\it  Avenida Jos\'{e} Mart\'{i} No. 270, Pinar del R\'{i}o, Cuba. CP 20100}\\

\vspace{.4cm}
{\it ${}^{a,b}$ The Abdus Salam International Centre for Theoretical Physics}\\
{\it Strada Costiera 11, 34014 Trieste, Italy}\\

\vspace{.4cm}
{\it ${}^{b}$  Leinweber Center for Theoretical Physics,  Department of Physics}\\
{\it University of Michigan, Ann Arbor, MI 48109, USA}\\
\vspace{.4cm}

\end{center}

\vskip 1 cm

\vskip 1.5 cm
\begin{abstract}
We study the superconfomal index of four-dimensional  toric quiver gauge theories using a Bethe Ansatz approach recently applied by Benini and Milan. Relying  on a particular set of solutions to the corresponding Bethe Ansatz equations we evaluate the superconformal index in the large $N$ limit, thus avoiding to take any Cardy-like limit. We present explicit results for theories arising as a stack of $N$ D3 branes at the tip of toric  Calabi-Yau cones: the conifold theory, the suspended pinch point gauge theory,  the first del Pezzo  theory and $Y^{p,q}$ quiver gauge theories. For a suitable choice of the chemical potentials of the theory we find agreement with predictions made for the same theories in the Cardy-like limit. However, for other regions of the domain of chemical potentials the superconformal index is modified and consequently the associated black hole entropy receives corrections. We work out explicitly the simple case of the conifold theory. 
\end{abstract}

\end{titlepage}


\section{Introduction}

The understanding of the quantum microstates responsible for the entropy of black holes has long been one of the central questions in the path to a quantum theory of gravity. In the context of the AdS/CFT correspondence it has recently  been shown that the entropy of certain asymptotically AdS$_4$ black holes admits a microscopic explanation in terms of a topologically twisted field theory \cite{Benini:2015eyy} (see  \cite{Hosseini:2018qsx,Zaffaroni:2019dhb} for reviews with extensive lists of references).

More recently, the question of microstates for asymptotically AdS$_5$ black holes dual to $\mathcal{N} =4$ supersymmetric Yang-Mills  (SYM), which was originally tackled in \cite{Kinney:2005ej}, has been revisited providing a microscopic entropy matching using various approaches. A broader interpretation of localization was successfully put forward in   \cite{Cabo-Bizet:2018ehj} while an analysis of the free-field partition function in a particular limit led to the entropy in \cite{Choi:2018hmj} (see also \cite{Choi:2018vbz}). Both these groups relied on a particular Cardy-like limit to evaluate the path integral. Another approach, put forward by Benini and Milan  in \cite{Benini:2018ywd}, attacked the superconformal index using a Bethe Ansatz approach developed in  \cite{Benini:2018mlo}.  Understanding that the superconformal index can be written as a sum over solutions to Bethe Ansatz equations was demonstrated in \cite{Closset:2017bse} based on  interesting relations between observables on manifolds of different topologies  developed in \cite{Closset:2017zgf}. One key advantage of the Bethe Ansatz approach is that it does not require taking the Cardy limit and thus opens the door for a more in-depth understanding of the superconformal index. In this brief note  we simply  generalize the  large $N$ results obtained for ${\cal N}=4$ SYM using the Bethe Ansatz approach to a large class of ${\cal N}=1$ 4d supersymmetric field theories.

Other recent studies demonstrating that the Cardy-like limit of the superconformal index of 4d ${\cal N} = 4 $ SYM accounts for the entropy function, whose Legendre transform corresponds to the entropy of the holographically  dual AdS$_5$ rotating black
holes were presented in \cite{ArabiArdehali:2019tdm,Honda:2019cio}.  Such analysis has by now been extended to generic  ${\cal N}=1$ supersymmetric gauge theories \cite{Cabo-Bizet:2019osg,Kim:2019yrz} including a particular description specialized to  arbitrary ${\cal N}=1$ toric quiver gauge theories, 
observing that the corresponding entropy function can be interpreted in terms of the toric data \cite{Amariti:2019mgp}. These powerful results rest on systematic studies of the Cardy limit developed in, for example, \cite{Ardehali:2014zba,DiPietro:2014bca,Ardehali:2015bla,Ardehali:2016kza,DiPietro:2016ond}.

In this note we verify that a class of holonomies  of the form $u_i- u_j=\frac{\tau}{N}(i -j)$,  used prominently in  \cite{Benini:2018ywd} for the case of   $\mathcal{N}=  4$ SYM,   can be  generalized to evaluate the superconformal index of generic ${\cal N}=1$ four-dimensional superconformal field theories.

The rest of the note is organized as follows. In section \ref{Sec:BA_SCI} we show that a particular class of holonomies  solves the Bethe Ansatz equation for generic 4d ${\cal N}=1$ gauge theories and proceed to evaluate the superconformal index in the large $N$ limit. Section \ref{Sec:Explicit} works out explicitly the index for a number of superconformal field theories. We find that there is always a way of redefining the chemical potentials suitably, such that the superconformal index obtained reproduces successfully the entropy of the dual $AdS_5$ black holes upon extremization of its Legendre transform.  We also focus on the conifold theory  in which the simplicity of the superconformal index allows us to study it for some region of the domain of chemical potentials that can provide a black hole entropy with corrections purely depending on the angular velocity $\tau$.   We conclude in section \ref{Sec:Conclusions}. \\ \par 
\textbf{Note added:} After this manuscript was originally submitted to the arxiv we received  \cite{Lanir:2019abx} with a considerable overlap with this work. The authors of \cite{Lanir:2019abx} perform a more exhaustive analysis of the behavior of the entropy function for different regions in the domain of complex chemical potentials. The present version of this manuscript contains substantial changes with respect to the first two versions appearing in arxiv. We have essentially found that, selecting a set of chemical potentials that ensures an optimal obstruction of cancellations between bosonic and fermionic contributions to the superconformal index, one can always find a region of chemical potentials where the index accounts for the black hole entropy. This resolves an apparent tension between our conclusions and the ones subsequently reported in \cite{Lanir:2019abx}.  
 
\section{Bethe Ansatz approach to the superconformal index}\label{Sec:BA_SCI}
In this section we generalize the solutions to the Bethe Ansatz type equations proposed in \cite{Closset:2017bse, Benini:2018ywd, Benini:2018mlo} to evaluate the superconformal index of ${\cal N}=4$ SYM to generic 4d ${\cal N}=1$ supersymmetric gauge theories. For concreteness we will work in the context of toric quiver gauge theories which are naturally decorated with extra global and baryonic symmetries but the results apply more generally to $4d$ ${\cal N}=1$ supersymmetric gauge theories.\\ \par
Consider a generic $\mathcal{N}=1$ theory with semi-simple gauge group $G$, flavor symmetry $G_F$ and non-anomalous $U\left(1\right)_R$ R-symmetry. The matter content of this theory is taken to be $n_{\chi}$ chiral multiplets $\Phi_a$ in representations $\mathfrak{R}_a$ of $G$, with flavor weights $\omega_a$ in some representation $\mathfrak{R}_F$ of $G_F$ and superconformal R-charge $r_a$. Let us start by introducing the following quantities which are related to global fugacities and holonomies in the Cartan of the gauge group:
\begin{eqnarray}
p =e^{2 \pi i  \tau}, \hspace{3mm} q =e^{2 \pi i \sigma}, \hspace{3mm} v_{\alpha} = e^{2 \pi i \xi_{\alpha}}, \hspace{3mm} z_i = e^{2 \pi i u_i} \label{fugacities}
\end{eqnarray}
and the R-charge chemical potential which is fixed by supersymmetry to:
\begin{eqnarray}
\nu_R = \frac{1}{2}\left(\tau + \sigma\right). \label{nur}
\end{eqnarray}
With the above data, the integral representation for the superconformal index can be written as  \cite{Romelsberger:2007ec,Assel:2014paa}:
\begin{eqnarray}
\mathcal{I} \left(p , q ; v\right) & = & \frac{\left(p ; p \right)_{\infty}^{\text{rk}\left(G\right)}\left(q ; q \right)_{\infty}^{\text{rk}\left(G\right)}}{|\mathcal{W}_G|} \oint_{\mathbb{T}^{\text{rk}\left(G\right)}}\frac{\prod_{a=1}^{n_{\chi}}\prod_{\rho_a \in \mathfrak{R}_a}\Gamma_e \left[\left(p q\right)^{r_a/2}z^{\rho_a}v^{\omega_a};p,q\right]}{\prod_{\alpha \in \Delta}\Gamma_e \left(z^{\alpha}; p, q\right)} \prod_{i =1}^{\text{rk}\left(G\right)}\frac{d z_i}{2 \pi i z_i}. \label{index}
\end{eqnarray}
The integration variables $z_i$ parameterize the maximal torus of the gauge group $G$ and the integration contour is the product of $\text{rk}\left(G\right)$  unit circles. Following standard notation, $\rho_a$ are the weights of the representation $\mathfrak{R}_a$, $\alpha$ parameterize the roots of $G$ and $|\mathcal{W}_G|$ is the order of the Weyl group. The notation adopted also denotes $z^{\rho_a} \equiv \prod_{i = 1}^{\text{rk}\left(G\right)} z_i^{\rho_a^i}$ and $v^{\omega_a} =\prod_{\alpha = 1}^{\text{rk}\left(G_F\right)} v_{\alpha}^{\omega_a^{\alpha}} $. The other functions involved in the expression for the superconformal index are the Elliptic Gamma function
\begin{eqnarray}
\Gamma_e \left(z; p, q\right)= \prod_{m,n =0}^{\infty} \frac{1 - p^{m+1}q^{n+1}/z}{1 - p^m q^{n}z}, \hspace{2mm}|p| <1 , \hspace{1.5mm} |q| <1, \label{GammaE}
\end{eqnarray}
and the q-Pochhamer symbol
\begin{eqnarray} 
\left(z;q\right)_{\infty}= \prod_{n=0}^{\infty}\left(1 - z q^n\right), \hspace{2mm} |q|<1. \label{Poch}
\end{eqnarray}
An interesting result of \cite{Benini:2018mlo} and \cite{Benini:2018ywd}, based on \cite{Closset:2017bse}, is to rewrite the above superconformal index in terms of solutions to certain Bethe Ansatz like system of equations taking the generic form of 
\begin{eqnarray}
Q_i \left(u ; \xi, \nu_{R}, \omega\right) =  1 \hspace{2mm} \forall \hspace{2mm} i = 1, ..., \text{rk}\left(G\right) \label{BAeq}
\end{eqnarray}
where $\omega$ is such that $r \tau = s \sigma$ with $r $ and  $s$ coprime integer numbers (in practice we will evaluate the equations for $r = s $). Furthermore, the ``Bethe Ansatz operator'' is defined as:
\begin{eqnarray}
Q_i \left(u ; \xi, \nu_{R}, \omega\right)& = & \prod_{a=1}^{n_{\chi}}\prod_{\rho_a \in \mathfrak{R}_a} P\left(\rho_a \left(u\right) +\omega_a \left(\xi \right) +r_a \nu_R ; \hspace{1mm}\omega \right)^{\rho_a^i},
\end{eqnarray} 
where 
\begin{eqnarray}
P\left(u; \omega\right) & = & \frac{e^{- \pi i \frac{u^2}{\omega}+\pi i u}}{\theta_0 \left(u; \omega\right)}. \\ \nonumber 
\end{eqnarray}
Thus, 
\begin{eqnarray}
P\left(\rho_a \left(u\right) +\omega_a \left(\xi \right) +r_a \nu_R ; \hspace{1mm}\omega \right)& = &\frac{e^{- \pi i \frac{1}{\omega}\left(\rho_a \left(u\right) +\omega_a \left(\xi \right) +r_a \nu_R  \right)^2+\pi i \left(\rho_a \left(u\right) +\omega_a \left(\xi \right) +r_a \nu_R \right)}}{\theta_0 \left(\rho_a \left(u\right) +\omega_a \left(\xi \right) +r_a \nu_R ; \omega\right)} ,\label{BAop}
\end{eqnarray}
where:
\begin{eqnarray}
\theta_0 \left(u; \omega\right) = \left(e^{2 \pi i u}; e^{2 \pi i \omega}\right)_{\infty} \left(e^{2 \pi i \left(\omega- u\right)}; e^{2 \pi i \omega}\right)_{\infty} .\label{theta0}
\end{eqnarray}
Now we would like to evaluate  the Bethe Ansatz equations for the case of a toric quiver gauge theory.  Toric quiver gauge theories describe the low energy dynamics of a stack of $N$ D3 branes probing the tip of a toric Calabi-Yau singularity;  there is by now a vast literature detailing how to construct a supersymmetric field theory given toric data (see, for example, \cite{Benvenuti:2004dy,Franco:2005sm}).  Consider a toric quiver gauge theory whose gauge group $G$ has $n_{\text{v}}$ simple factors (in all the $\mathcal{N} = 1$ quiver gauge theories we will deal with, the number of simple factors coincides with the number of vector multiplets). We focus, for concreteness, on the case in which all the gauge group factors are $SU(N_a)$, $a$ goes from $1$ to $n_{\text{v}}$, with $N_a=N \hspace{2mm} \forall \hspace{1mm} a$, the same numerical value for all nodes. In these theories the weight vectors $\rho$ are such that for any bi-fundamental field $\Phi_{ab}$ (notice that in the more generic notation used in \cite{Benini:2018mlo}, the index $a$ of $\Phi_a$ would now split into $ab$):
\begin{eqnarray}
\rho_{ij}^{\Phi_{ab}}\left(u\right) \equiv u_{ij}^{ab} \equiv u_i^{a} - u_j^b.
\end{eqnarray} 
Let us now evaluate the operator $P(u;\omega)$ for a generic field $\Phi_{ab}$ (when $\Phi_{ab}$ transforms in the adjoint representation of $G$ then, in this notation, $a = b$):
\begin{eqnarray}
Q_{i_a} \left(u ; \xi, \tau,\sigma, \omega\right)=\prod_{(a,b)}\prod_{j_b}\prod_{\rho_{ij}^{(a,b)}}P\left( u_{i_a} - u_{j_b} + \sum_{l=1}^{d-1}q_{(a,b)}^l \Delta_l + r_{ab}\nu_R\right)^{\rho_{ij}^{(a,b)}},
\end{eqnarray}    
where $(a,b)$ run over all the fields $\Phi_{ab}$ for a fixed $a$ and $r_{ab}$ are the $R-$charges of the fields $\Phi_{ab}$. The $d - 1$ fugacities correspond to the flavor symmetries appearing in the generic toric gauge theories that we will study,  $d$ is the number of external points of the toric diagram that are related to the quivers defining the theory \cite{Amariti:2019mgp}.  If we denote $\langle a,b\rangle \equiv (a,b)|_{\rho_{ij}^{(a,b)}>0}$, which implies:
\begin{eqnarray}
Q_{i_a} \left(u ; \xi, \tau,\sigma, \omega\right)& = &\prod_{\langle a,b \rangle}\prod_{j_b}\prod_{\rho_{ij}^{\langle a,b \rangle}}\left[\frac{P\left( u_{i_a} - u_{j_b} + \sum_{l=1}^{d-1}q_{\langle a,b \rangle}^l \Delta_l + r_{ab}\nu_R\right)}{P\left( u_{j_b} - u_{i_a} + \sum_{l=1}^{d-1}q_{\langle b,a \rangle}^l \Delta_l  + r_{ba}\nu_R\right)}\right]^{\rho_{ij}^{\langle a,b \rangle}} \label{long} \\ \nonumber
& = & \prod_{\langle a,b \rangle}\prod_{j_b}\prod_{\rho_{ij}^{\langle a,b \rangle}}\left[\frac{e^{-2\pi i \left( -u_{i_a} + u_{j_b}\right)}\theta_0\left( -u_{j_b} + u_{i_a} + \sum_{l=1}^{d-1}q_{\langle a,b \rangle}^l \Delta_l + r_{ab}\nu_R ; \omega\right)}{\theta_0\left( u_{i_a} - u_{j_b} + \sum_{l=1}^{d-1}q_{\langle b,a \rangle}^l \Delta_l  + r_{ba}\nu_R ; \omega \right)}\right]^{\rho_{ij}^{\langle a,b \rangle}}  \\ \nonumber
& = & e^{-2\pi i \sum_{j_b}\left( u_{i_a} - u_{j_b}\right)}\prod_{\langle a,b \rangle}\prod_{j_b}\frac{\theta_0\left( -u_{i_a} + u_{j_b} + \sum_{l=1}^{d-1}q_{\langle a,b \rangle}^l \Delta_l + r_{ab}\nu_R; \omega\right)}{\theta_0\left( -u_{j_b} + u_{i_a} + \sum_{l=1}^{d-1}q_{\langle b,a \rangle}^l \Delta_l + r_{ab}\nu_R ; \omega\right)}. 
\end{eqnarray}
 Let us now introduce a Lagrange multiplier $\lambda_a$ that accounts for the constraint ensuring the condition $\sum_{i}u_i^a = 0$ \cite{Benini:2018ywd},  with its help, equation (\ref{long}) can be written as:
\begin{eqnarray}
Q_{i_a} \left(u ; \xi, \tau,\sigma, \omega\right)& = & e^{2\pi i\left( \sum_b\lambda_b - \sum_{j_b} u_{ij}^{ab}\right)}\prod_{\langle a,b \rangle}\prod_{j_b}\frac{\theta_0\left( -u_{ij}^{ab} + \sum_{l=1}^{d-1}q_{\langle a,b \rangle}^l \Delta_l + r_{ab}\nu_R  ; \omega\right)}{\theta_0\left( -u_{ji}^{ba}  + \sum_{l=1}^{d-1}q_{\langle b,a \rangle}^l \Delta_l + r_{ba}\nu_R  ; \omega\right)}, \label{baeq1}
\end{eqnarray} 
where we have denoted $u_{i_a} - u_{j_b} \equiv u_{ij}^{ab}$. Restricting ourselves to the case with $\tau = \sigma$, we would like to propose a set of $u_{ij}^{ab}$ that makes (\ref{baeq1}) equal to $1$, thus solving the Bethe Ansatz equation (\ref{BAeq}). It is natural to make an attempt with a direct generalization of the type of solution encountered in \cite{Benini:2018ywd}, namely: $u_{ij}^{ab} = \frac{\tau}{N}\left(i_a - j_b\right)$. These solutions appeared first in \cite{Hosseini:2016cyf} while evaluating the topologically twisted of 4d ${\cal N}=1$ theories on $T^2\times S^2$ in the high temperature limit; it was later shown in \cite{Hong:2018viz} that such configuration provides an exact solution to the Bethe Ansatz equations.

Consider one generic factor entering in (\ref{baeq1}) for a fixed value of $b$:
\begin{eqnarray}
\prod_{j_b} \frac{\theta_0\left( u_{ij}^{ab} + \Delta_{ab}; \omega\right)}{\theta_0\left( -u_{ij}^{ab}  + \Delta_{ba} ; \omega\right)}\big{|}_{u_{ij}^{ab} = \frac{\tau}{N}\left(i_a - i_b\right)} & = & \frac{\prod_{k=0}^{i_a-1}\theta_0\left(\frac{\tau}{N}k +\Delta_{ab}\right) \times\prod_{k = i_a - N}^{-1} \theta_0 \left(\frac{\tau}{N}k + \Delta_{ab}\right)}{\prod_{k=0}^{N- i_a}\theta_0 \left(\frac{\tau}{N}k + \Delta_{ba}\right) \times \prod_{k =i_a -1}^{-1} \theta_0 \left(\frac{\tau}{N} k +\Delta_{ba}\right)} \label{this} \\ \nonumber
\text{with} \hspace{2mm}\Delta_{ab} & \equiv &  \sum_{l=1}^{d-1}q_{\langle a,b \rangle}^l \Delta_l + r_{ab}\tau\\ \nonumber
\prod_{j_b} \frac{\theta_0\left( u_{ij}^{ab} + \Delta_{ab}; \omega\right)}{\theta_0\left( -u_{ij}^{ab}  + \Delta_{ba} ; \omega\right)} & = & \frac{\prod_{k=0}^{N-1}\theta_0 \left(\frac{\tau}{N}k + \Delta_{ab}\right) \times \prod_{k=i_a -N}^{-1}\left(- e^{2 \pi i \tau \frac{k}{N}}e^{2 \pi i \Delta_{ab}}\right)}{\prod_{k=0}^{N-1}\theta_0 \left(\frac{\tau}{N}k + \Delta_{ba}\right) \times \prod_{k =1 - i_a}^{-1}\left(- e^{2 \pi i \tau \frac{k}{N}}e^{2 \pi i \Delta_{ba}}\right)} \\ \nonumber
& =&  \frac{\prod_{k=0}^{N-1}\theta_0 \left(\frac{\tau}{N}k + \Delta_{ab}\right)}{\prod_{k=0}^{N-1}\theta_0 \left(\frac{\tau}{N}k +\Delta_{ba}\right)}\times \left( e^{\pi i (-1+\tau)}\right)^{\left(2 i_a - N - 1\right)}e^{- 2 \pi i \left[i_a\left(\Delta_{ab}+\Delta_{ba}\right) - N \Delta_{ab} - \Delta_{ba}\right]} \\ \nonumber
&= & \frac{\prod_{k=0}^{N-1}\theta_0 \left(\frac{\tau}{N}k + \Delta_{ab}\right)}{\prod_{k=0}^{N-1}\theta_0 \left(\frac{\tau}{N}k +\Delta_{ba}\right)}\times e^{2 \pi i_a \left(\tau - 1 - \Delta_{ba} - \Delta_{ab}\right)}e
^{\pi i\left[(1- \tau)(1+N)+ 2(\Delta_{ba}+ N \Delta_{ab})\right]} \\ \nonumber
& \equiv &  F(\Delta_{ab}, \Delta_{ba}, \tau) e^{2 \pi i_a \left(\tau - 1 - \Delta_{ba} - \Delta_{ab}\right)} 
\end{eqnarray}
In (\ref{this}) we have used the following properties of the $\theta_0$ function:
\begin{eqnarray}
\theta_0 \left(u +n + m \tau; \tau \right) & = &  - e^{-2 \pi m i u - \pi i m \tau (m-1)}\theta_0 \left(u; \tau\right) \\ \nonumber
\theta_0\left(u ; \tau\right) & = & \theta_0 \left(\tau - u; \tau\right) =  - e^{2 \pi i u}\theta_0 \left(- u; \tau\right),
\end{eqnarray}
and for the sake of compactness we have absorbed all the factors independent of $i_a$ in the function $F(\Delta_{ab},\Delta_{ba} ,\tau)$.  Inserting (\ref{this}) back into  (\ref{baeq1}) leads to multiplying all the results obtained in (\ref{this}) for all  $n_a$ values of $b$ connected with $a$ via some field $\Phi_{ab}$ :
\begin{eqnarray}
Q_{i_a} \left(u ; \xi, \tau\right) & = &e^{2\pi i\left( \sum_{b}\lambda_b - \sum_{j_b}\frac{\tau}{N}\left(i_a - j_b\right)\right)}  F_a(\tau) e^{2 \pi i_a \left[n_a(\tau - 1) -\sum_{b=1}^{n_a} (\Delta_{ba} + \Delta_{ab})\right]}\\ \nonumber
\text{where} \hspace{1mm} F_a(\tau)& \equiv & \prod_{b=1}^{n_a} F(\Delta_{ab}, \Delta_{ba}, \tau) \\ \nonumber
\sum_{b=1}^{n_a}\left(\Delta_{ab} + \Delta_{ba}\right)& = &  \sum_{b=1}^{n_a} r_{ab}\tau  \\ \nonumber
& \Downarrow & \\ \nonumber
Q_{i_a} \left(u ; \xi, \tau\right)& = & e^{2\pi i\left(  \sum_{b}\lambda_b - n_a\frac{\tau}{N}\left(N i_a -\frac{ N(N-1)}{2}\right)\right)}F_a(\tau) e^{2 \pi i_a n_a(\tau - 1)}  \\ \nonumber
& = & e^{2\pi i\left( \sum_{b}\lambda_b - n_a \tau \frac{N-1}{2}\right)} F_a(\tau) e^{-2 \pi i_a n_a} .
\end{eqnarray}
Upon a proper choice for the Lagrange multipliers we can ensure that:
\begin{eqnarray}
Q_{i_a} \left(u ; \xi, \tau\right)& = & e^{-2 \pi i_a n_a} =1 \blacksquare.
\end{eqnarray}

\subsection{Evaluation of the index}
The formula for the superconformal index in terms of solutions to the Bethe Ansatz like equations reads    \cite{Closset:2017bse,Benini:2018mlo}:
\begin{eqnarray}
\mathcal{I}\left(p,q;v\right) & = & \kappa_{G}\sum_{\hat{u} \in \mathfrak{M}_{BAE}}\mathcal{Z}_{tot} \left(\hat{u};\xi, \nu_R,  r \omega, s \omega \right) H \left(\hat{u}; ,\xi, \nu_R, \omega\right)^{-1} \label{for} \\ \nonumber
\kappa_{G} & = &  \frac{\left(p ; p \right)_{\infty}^{\text{rk}\left(G\right)}\left(q ; q \right)_{\infty}^{\text{rk}\left(G\right)}}{|\mathcal{W}_G|}\\ \nonumber
\mathcal{Z}_{tot}\left(u;\xi, \nu_R,  r \omega, s \omega \right) & = & \sum_{\left\{m_{i_a}\right\} =1}^{rs} \mathcal{Z}\left(u - m \omega ;\xi, \nu_R,  r \omega, s \omega \right) \\ \nonumber
\mathcal{Z}\left(u  ;\xi, \nu_R, r\omega, s \omega \right) & = & \frac{\prod_{\Phi_{ab}}\prod_{i_a\neq j_b}\Gamma_e \left( u_{i_a} - u_{j_b} + \Delta_{ab}; \tau, \sigma \right)}{\prod_{\alpha \in \Delta}\Gamma_e \left(\alpha \left(u\right);\tau,\sigma\right)}
\\\nonumber
H \left(u; \xi, \nu_R, \omega\right) & = & \text{det}\left[\frac{1}{2 \pi i} \frac{\partial Q_{i_a}\left(u; ,\xi, \nu_R, \omega\right)}{\partial u_{j_b}}\right]_{i_a j_b}.
\end{eqnarray}
We assume that dominant contributions to the index in the large $N$ limit will come from terms analogous to those  dominating the expression obtained in \cite{Benini:2018ywd} for the $\mathcal{N}= 4$ SYM theory. This implies that in order to investigate the large $N$ limit of (\ref{for}), we only need to consider the following term:
\begin{eqnarray}
\Gamma_e \left(u_{ij}^{ab} + \Delta_{ab}; \tau, \tau\right) & = & \frac{e^{- \pi i \mathcal{Q}\left(u_{ij}^{ab} + \Delta_{ab};\tau, \tau\right)}}{\theta_0 \left(\frac{u_{ij}^{ab} + \Delta_{ab}}{\tau}; - \frac{1}{\tau}\right)} \times \prod_{k =0}^{\infty}
\frac{\psi \left(\frac{k +1 +u_{ij}^{ab}}{\tau}\right)}{\psi \left(\frac{k - u_{ij}^{ab} - \Delta_{ab}}{\tau}\right)} \label{EGF} \\ \nonumber
\mathcal{Q} \left(u; \tau, \sigma\right) & = & \frac{u^3}{3 \tau \sigma} - \frac{\tau + \sigma -1}{2 \tau \sigma} u^2 + \frac{\left(\tau + \sigma\right)^2 + \tau \sigma - 3\left(\tau + \sigma\right) +1}{6 \tau \sigma}u + \frac{\left(\tau+ \sigma - 1\right)\left(\tau +\sigma - \tau \sigma\right)}{12 \tau \sigma} \\ \nonumber
\mathcal{Q} \left(u + \Delta; \tau, \tau\right) & = & \frac{u^3}{3 \tau^2} + u^2 \left(\frac{\Delta}{\tau^2} - \frac{2 \tau -1}{2 \tau^2} \right) + u \left(\frac{1 - 6 \tau + 5 \tau^2}{6 \tau^2} + \frac{\Delta^2}{\tau^2} - \frac{2 \tau -1}{ \tau^2} \Delta\right) - \\ \nonumber
& - & \frac{\Delta^2}{2 \tau^2}\left(2 \tau -1\right)+ \frac{\Delta}{6 \tau^2} \left(5 \tau^2 - 6 \tau + 1\right) + \frac{1}{12 \tau^2} \left(2 \tau - 1\right)\left(2 \tau - \tau^2\right) + \frac{\Delta^3}{\tau^2}.
\end{eqnarray}
Note that, the leading contribution coming from the vector multiplets can be obtained from  (\ref{EGF}) by setting $\Delta_{ab} = 0$. In the large $N$ limit we can write:
\begin{eqnarray}
\log \mathcal{I}\big{|}_{\text{large}\hspace{1mm}N} & =& \sum_{\Phi_{ab}}\sum_{i_a, j_b}\log \Gamma_e \left(u_{ij}^{ab}+ \Delta_{ab}; \tau, \tau\right) \big{|}_{\text{large}\hspace{1mm}N}-\sum_{\alpha \in \Delta }\Gamma_e \left(\alpha \cdot u, \tau, \tau\right)\big{|}_{\text{large}\hspace{1mm}N} \label{logi0} .
\end{eqnarray}
As a clarifying example, let us now analyze the case of $\mathcal{N} = 4$ SYM theory already studied in \cite{Benini:2018ywd} and peroform the same calculation using the toric data language of  \cite{Amariti:2019mgp}. The corresponding $\Phi_{ab}$ are the three chiral fields $\Phi_{1,2,3}$ appearing in the superpotential :
\begin{eqnarray}
W = \text{Tr} \left(\Phi_{1}\left[\Phi_{2},\Phi_3\right] \right),
\end{eqnarray}
with the associated chemical potentials being $\Delta_{1,2,3}$. Accordinng to our definition of the chemical potentials we have that, for the $R-$charge assignment used in  \cite{Benini:2018ywd}:
\begin{eqnarray}
\Delta_{\Phi_1} & = & \Delta_1 \label{deltaI} \\ \nonumber
\Delta_{\Phi_2} & = & \Delta_2 \\ \nonumber
\Delta_{\Phi_3}  &= & 2 \tau - \Delta_1 - \Delta_2
\end{eqnarray}
Using the identity:
\begin{eqnarray}
\Gamma_e \left(\Delta + 2 \tau ;  \tau, \tau\right) =  \frac{1}{\Gamma_e \left(- \Delta; \tau , \tau\right)} \label{identity}
\end{eqnarray}
reduces (\ref{logi0}) to the following expression:
\begin{eqnarray}
\log \mathcal{I}\big{|}_{\text{Large} \hspace{1mm} N } & = &  \sum_{i,j } \log \Gamma_e \left(u_{ij}^{1}+\Delta_{1} ;  \tau, \tau\right)\big{|}_{\text{Large} \hspace{1mm} N } +\log \Gamma_e \left(u_{ij}^{2}+\Delta_{2} ;  \tau, \tau\right)\big{|}_{\text{Large} \hspace{1mm} N } + \label{123} \\ \nonumber
 & - & \log \Gamma_e \left(u_{ij}^{3}+\Delta_{1} + \Delta_2 ;  \tau, \tau\right)\big{|}_{\text{Large} \hspace{1mm} N }   - \frac{i  \pi N^2}{3\tau^2}\tau\left( \tau - \frac{1}{2}\right)\left( \tau -1\right) \\ \nonumber
& = & - \frac{i  \pi N^2}{3\tau^2} \left(\left[\Delta_{1}\right]_{\tau} -\tau\right)\left(\left[\Delta_{1}\right]_{\tau}- \tau + \frac{1}{2}\right)\left(\left[\Delta_{1}\right]_{\tau}- \tau +1\right) - \\ \nonumber
& - & \frac{i  \pi N^2}{3\tau^2} \left(\left[\Delta_{2}\right]_{\tau} -\tau\right)\left(\left[\Delta_{2}\right]_{\tau}- \tau + \frac{1}{2}\right)\left(\left[\Delta_{2}\right]_{\tau}- \tau +1\right) +  \\ \nonumber
&+&  \frac{i  \pi N^2}{3\tau^2} \left(\left[\Delta_{1}+ \Delta_2\right]_{\tau} -\tau\right)\left(\left[\Delta_{1} + \Delta_2\right]_{\tau}- \tau + \frac{1}{2}\right)\left(\left[\Delta_{1}+ \Delta_2\right]_{\tau}- \tau +1\right) - \\ \nonumber
& - & \frac{i  \pi N^2}{3\tau^2}\tau\left( \tau - \frac{1}{2}\right)\left( \tau -1\right).
\end{eqnarray}
where  $\left[\Delta_{ab}\right]_{\tau}$ is defined such that $\left[\Delta\right]_{\tau} = \Delta \hspace{2mm} \text{mod} \hspace{2mm} 1$ \cite{Benini:2018ywd} and depends on the region withing the domain of complex chemical potentials one is evaluating (for a more detailed description of this function see also \cite{Lanir:2019abx}).
If $|\Delta_{ab}|<1$ ,  then :
\begin{eqnarray}
\log \mathcal{I}\big{|}_{\text{Large} \hspace{1mm} N } & = & - \frac{i \pi N^2}{\tau^2}\Delta_1 \Delta_2 \left(2 \tau - \Delta_1- \Delta_2 \right), \label{resultn4}
\end{eqnarray}
which is indeed the necessary structure in order for the superconformal index of $\mathcal{N}=4$ SYM to account for the entropy of the dual AdS$_5$ black hole \cite{Benini:2018ywd}. \\ \par 

Before proceeding to generic toric quiver gauge theories, let us comment on the choice of $R$-charge assignment, since one might expect a more symmetric one based on $a$-maximization. We notice that, if one chooses a set of chemical potentials and $R-$charges as the one used in \cite{Amariti:2019mgp}, namely where $r_1 = r_2 = r_3 = \frac{2}{3}$, in contrast with the choice $r_1 = r_2 = 0, \hspace{1.5mm} r_3 = 2$, then the use of identity (\ref{identity}) is not directly possible. This means that, if one starts with the data suggested by $a-$maximization \cite{Amariti:2019mgp} $(r_1 = r_2 = r_3 = \frac{2}{3})$, then (\ref{resultn4}) should be understood in terms of shifted chemical potentials that would permit some of the arguments of the Elliptic Gamma functions in  (\ref{logi0}) to have the structure $\Delta + 2 \tau$ as needed in (\ref{identity}). Specifically ,  we have:
\begin{eqnarray}
\Delta_{\Phi_1} + \frac{2}{3} \tau & =& \Delta_1 + \frac{2}{3} \tau = \left(\Delta_1 + \frac{2}{3}\tau\right) + \frac{2}{3}\tau - \frac{2}{3} \tau \rightarrow \Delta_1 \label{shiftN4} \\ \nonumber
\Delta_{\Phi_2} + \frac{2}{3}\tau & = &  \Delta_2 + \frac{2}{3} \tau = \left(\Delta_2 + \frac{2}{3}\tau\right) + \frac{2}{3}\tau - \frac{2}{3} \tau \rightarrow \Delta_2 \\ \nonumber
\Delta_{\Phi_3}+ \frac{2}{3} \tau & = &  - \Delta_1 - \Delta_2 + \frac{2}{3}\tau \rightarrow  - \Delta_1 - \Delta_2 + 2 \tau ,
\end{eqnarray}

We can either interpret this as a suitable redefinition of the chemical potentials wich does not affect the physical $R-$charge obtained via $a-$maximization or rather as a computation done directly with the more naive $R-$charge assignment used in \cite{Benini:2018ywd}. Let us now explore more generically the consequences of shifting  $\Delta_{ab}$ in such a way that the arguments of the elliptic Gamma functions in (\ref{logi0}) look either like $\Delta$ or $\Delta + 2 \tau.$   Suppose we do such a shift obtaining that a certain number, let us call this number $n_{s}$, of the total of $n_{\chi}$ chiral fields contributions to (\ref{logi0}) are of the form $ \Delta + 2 \tau$.  Thus, the leading contribution in $N$ to $\log \mathcal{I}$ takes the form:
\begin{eqnarray}
\log \mathcal{I} & = & - \frac{i  \pi N^2}{3\tau^2} \sum_{\Phi_{ab}} s_{ab}\left(\left[s_{ab}\Delta_{ab}\right]_{\tau} -\tau\right)\left(\left[ s_{ab}\Delta_{ab}\right]_{\tau}- \tau + \frac{1}{2}\right)\left(\left[s_{ab}\Delta_{ab}\right]_{\tau}- \tau +1\right)  -  \label{logi} \\ \nonumber
& - &\frac{i  \pi N^2}{3\tau^2}\sum_{\text{v}}\tau\left( \tau - \frac{1}{2}\right)\left( \tau -1\right) \\ \nonumber
& = &- \frac{i  \pi N^2}{3\tau^2} \sum_{\Phi_{ab}}  s_{ab}\left[ \left[s_{ab}\Delta_{ab}\right]_{\tau}\left(\left[s_{ab}\Delta_{ab}\right]_{\tau} + \frac{1}{2}\right)\left(\left[ s_{ab}\Delta_{ab}\right]_{\tau}+1\right)- 3 \tau \left[s_{ab}\Delta_{ab}\right]^2_{\tau}\right]] - \\ \nonumber
&  - & \frac{i  \pi N^2}{3\tau^2} \sum_{\Phi_{ab}}  s_{ab} \left[3 \tau^2\left[s_{ab}\Delta_{ab}\right]_{\tau} - 3 \tau \left[ s_{ab}\Delta_{ab}\right]_{\tau}\right] +\\ \nonumber
& + &\frac{i  \pi N^2}{3\tau^2}\left(n_{\chi}  - 2 n_s - n_{\text{v}}\right)\tau\left( \tau - \frac{1}{2}\right)\left( \tau -1\right),
\end{eqnarray}
where $\left[\Delta_{ab}\right]_{\tau}$ is defined such that $\left[\Delta\right]_{\tau} = \Delta \hspace{2mm} \text{mod} \hspace{2mm} 1$ \cite{Benini:2018ywd}, the sum $\sum_{\text{v}}$ is carried over the $n_{\text{v}}$ vector multiplets and $n_{\chi}$ is the number of chiral fields  , $s_{ab}$ is $1$ if $\Phi_{ab}$ effectively has $R-$charge $0$ and $-1$ if it has $R-$ charge $2$ with a new set of chemical potentials. Conservation of $U(1)$ charges implies $\sum_{\Phi_{ab}}\left[\Delta_{ab}\right]_{\tau} = 0$, which allows us to eliminate every linear term in $\left[\Delta_{ab}\right]_{\tau}$ appearing in (\ref{logi}), therefore we can write:
\begin{eqnarray}
\log \mathcal{I} & = &- \frac{i  \pi N^2}{3\tau^2} \sum_{\Phi_{ab}}s_{ab}K\left(s_{ab}\left[\Delta_{ab}\right]_{\tau}, \tau\right) + \label{similar0}\\ \nonumber
& + &\frac{i  \pi N^2}{3\tau^2}\left(n_{\chi}  - 2  n_s- n_{\text{v}}\right)\tau\left( \tau - \frac{1}{2}\right)\left( \tau -1\right).
\end{eqnarray}
where we have defined
\begin{eqnarray}
 K(\Delta , \tau) & \equiv &\left[\Delta\right]_{\tau}\left(\left[\Delta\right]_{\tau} + \frac{1}{2}\right)\left(\left[\Delta\right]_{\tau}+1\right) - 3 \tau \left[\Delta \right]^2_{\tau}  =  \frac{1}{2}\left(2 \Delta^3 - 3 |\Delta | \Delta +\Delta - 6 \tau |\Delta | \Delta \right)  \label{KK}
\end{eqnarray}
Recalling that:
\begin{eqnarray}
\left[\Delta + 1 \right]_{\tau} & = &\left[\Delta\right]_{\tau}, \\ \nonumber
\left[-\Delta\right]_{\tau} & = &  - \left[\Delta\right]_{\tau} -1 \\ \nonumber
\left[\Delta + \tau \right]_{\tau} & = & \left[\Delta\right]_{\tau}   + \tau.
\end{eqnarray} 
then (\ref{KK}) holds when $|\Delta | < 1$.  
\\ \par

Let us now analyze the properties of the function we have obtained. 
 Equation (\ref{similar0}) is very similar to the one obtained in \cite{Amariti:2019mgp} when analyzed in the Cardy-like limit of the index, however, there is an extra contribution of the form $\frac{i  \pi N^2}{3\tau^2}\left(n_{\chi} -2n_s- n_{\text{v}}\right)\tau\left( \tau - \frac{1}{2}\right)\left( \tau -1\right)$ which is still of order ${\cal O}(N^2)$ but sub-leading when $\tau \rightarrow 0$. Notice that at this point there is no dependence on  the holonomies of the gauge groups since we have already evaluated in the solutions of the Bethe Ansatz equations.  We still need to determine if we can find a consistent way of redefining the chemical potentials, thus fixing the value of $n_s$ and $s_{ab}$. The shifting has to preserve the $R-$ charge of the superpotential which is ensured by the constrain:
\begin{eqnarray}
\sum_{(ab)\in A} \Delta_{ab} = 2 \tau. \label{conns}
\end{eqnarray} 
where $A$ denotes monomial terms of the superpotential $W$. Let us call $n_F$ the number of elements in $A$.
 Using the fact that for these toric quiver gauge theories each chiral field appears only once in exactly two terms in the superpotential, then (\ref{conns}) implies that $2 n_s = n_{F}$.

To gain a better understanding of the implications that shifting the chemical potentials has on the  superconformal index of a toric quiver
gauge theory let us consider:
\begin{eqnarray}
\mathcal{I} & = & \text{Tr}_{BPS}\left(-1\right)^Fe^{- \beta H}e^{2 \pi i 2 \tau \left(J +\frac{1}{d}\sum_{I=1}^d Q_d \right)} e^{2 \pi i \sum_{i=1}^{d-1}\Delta_i\left(Q_i - Q_d\right)} \label{eq.Amariti},
\end{eqnarray}
where we have used the same basis for the non R-global symmetries used in  \cite{Amariti:2019mgp}. Shifting the chemical potentials as 
\begin{eqnarray}
\Delta_i \rightarrow \Delta_i - \frac{2 \tau}{d} \hspace{1.5mm} \text{ with} \hspace{1.5mm} i= 1, \cdots, d-1 \label{sh}
\end{eqnarray}
  allows us to rewrite the index as 
\begin{eqnarray}
\mathcal{I} & = & \text{Tr}_{BPS}\left(-1\right)^F e^{2 \pi i\left(\sum_{I=1}^d\Delta_I -2 \tau\right)Q_d}e^{- \beta H}e^{2 \pi i 2 \tau J } e^{2 \pi i \sum_{i=1}^{d-1}\Delta_i Q_i }\label{obstruction} \label{eq.Amariti}.
\end{eqnarray}
Exploiting the  constraint  
\begin{eqnarray}
\sum_{I =1}^d \Delta_I - 2 \tau  =   \pm 1 \label{constraint1}
\end{eqnarray}
 and identifying $e^{2 \pi i Q_d} = (-1)^F$ \cite{Amariti:2019mgp} allows us to express the superconformal index in such a way that bosonic-fermionic cancellations are optimally obstructed \cite{Choi:2018hmj}:
 \begin{eqnarray}
\mathcal{I} & = & \text{Tr}_{BPS}e^{- \beta H}e^{2 \pi i 2 \tau J } e^{2 \pi i \sum_{i=1}^{d-1}\Delta_i Q_i }\label{obstruction} \label{eq.Am2}.
\end{eqnarray}
 \
  
  The shifting \eqref{sh} which is dictated by the geometry of the toric diagram, in particular by its number of vertices, turns out to be the adequate one in order to reproduce the dual black hole entropy. \\ \par 
Finally, recalling that we are dealing with toric quivers, which can be drawn on a torus providing a polygonalization of the torus \cite{Benvenuti:2005ja} , and $n_{\chi}$ is the number of edges $n_E$ of the graph, $n_{F}$ is associated to the number of faces  and $n_{\text{v}}$ to the number of vertices then, the last term in (\ref{KK}) vanishes due to the Euler relation, 
$n_E - n_F - n_{\text{v}} =  0$ :
\begin{eqnarray}
\log \mathcal{I} & = &- \frac{i  \pi N^2}{3\tau^2} \sum_{\Phi_{ab}} s_{ab} K\left( s_{ab}\Delta_{ab}, \tau\right) +\frac{i  \pi N^2}{3\tau^2}\left(n_{\chi} - n_F - n_{\text{v}}\right)\tau\left( \tau - \frac{1}{2}\right)\left( \tau -1\right)\label{similar} \\ \nonumber 
& =&- \frac{i  \pi N^2}{3\tau^2} \sum_{\Phi_{ab}} s_{ab}  K\left(s_{ab}\Delta_{ab}, \tau\right). 
\end{eqnarray}
Defining $\Delta_d$ such that: $\sum_{I =1}^d \Delta_I - 2 \tau  =   - 1$ \cite{Benini:2018ywd}, it can be shown  that $\log \mathcal{I}$ can be writen as:
\begin{eqnarray}
\label{similar1}
\log \mathcal{I} & = &- \frac{i  \pi N^2}{6\tau^2} C_{I J K}\Delta_I \Delta_J \Delta_K \label{resultS} .
\end{eqnarray}
The coefficients $C_{IJK}$ in (\ref{similar1}) correspond, as pointed out originally in   \cite{Hosseini:2018dob} and later in  \cite{Amariti:2019mgp}, to the Chern-Simons couplings of the holographic dual gravitational description as elucidated in \cite{Benvenuti:2006xg}.  In the following section we proceed to evaluate the superconformal index for various  models, some of them recently discussed in a similar context in \cite{Amariti:2019mgp}, and compare our results with (\ref{similar1}).

\section{The superconformal index of various SCFT's }\label{Sec:Explicit}
\begin{figure}[H]
\caption{The figure shows the complex plane of chemical potentials for a generic $\Delta$ where the region specified by (\ref{const}) is shown in grey. } 
\centering
    \includegraphics[width=1\textwidth]{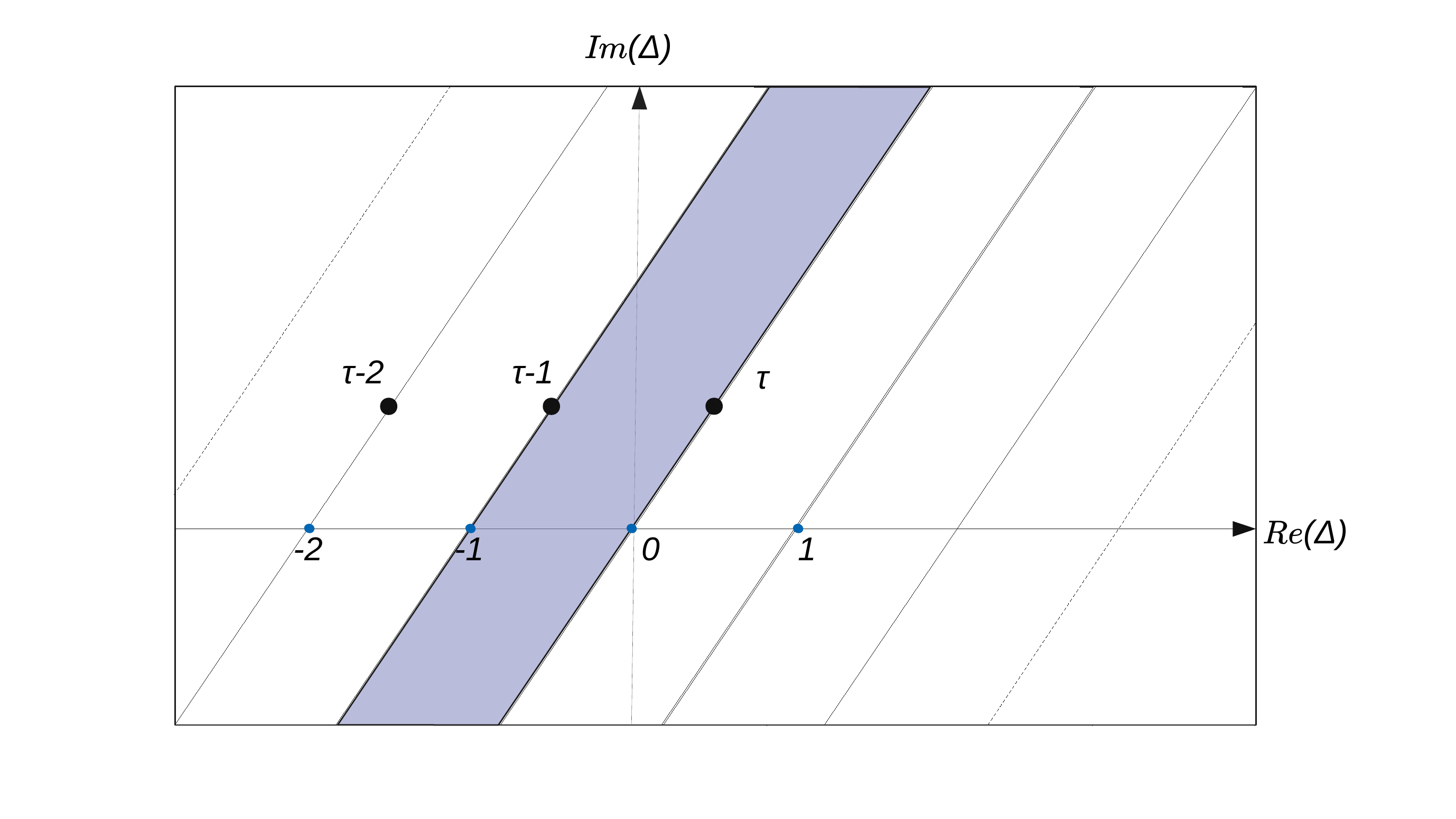}
    \label{gregion0}
    \end{figure}

We will apply our general result (\ref{similar1}) in various cases in each of which we follow the prescription of charge assignment used in \cite{Amariti:2019mgp}.  Indeed, below we will see that in order to obtain (\ref{resultS}) all the chemical potentials have to be shifted by $-\frac{2 \tau}{d}$, exactly like \cite{Amariti:2019mgp}.  We will restrict ourselves to the regime of chemical potentials $\Delta_i$ of the $d-1 \hspace{2mm} U(1)$ global symmetries such that:

\begin{eqnarray}
0 \leq |\Delta_i|  \leq \frac{1}{2} \hspace{2mm}\forall i, \hspace{2mm} 0 \leq \sum_{i =1}^{d-1}|\Delta_i |,
 \leq 1 \label{const1},
 \end{eqnarray} 

 which is inside the fundamental domain:

 \begin{eqnarray}
 \text{Im}\left(-\frac{1}{\tau}\right) & > &\text{Im}\left(\frac{\sum_{i =1}^{d-1}\left[\Delta_i\right]_{\tau}}{\tau}\right)>0\label{const},
 \end{eqnarray}

 which in our case will be useful to evaluate the function $K(\Delta, \tau)$ using equation (\ref{KK}). The region (\ref{const}) has been highlighted in Fig.\ref{gregion0} in grey. This regime also coincides with the one in which  the existence of a universal saddle point in which all the holonomies vanish according to the analysis carried in \cite{Amariti:2019mgp}, can be ensured.

\subsection{The conifold theory}\label{conifold}
We would like to study the index in the large $N$ limit and thus investigate it beyond the Cardy-like limit. To do so we start with  one of the simplest examples of toric quiver gauge theories -- the conifold theory \cite{Klebanov:1998hh} whose  quiver diagram is given below. We take the ranks of all the gauge groups equal ($N_1 = N_2 =N$) and the sub-index in $N_i$ helps describe the representations of the matter fields:\\ \par 
\begin{center} 
\begin{tikzpicture}
\node[state] at (0,0) (q2) {$N_1$};
\node[state, right of=q2] at (2,0) (q3) {$N_2$};
\draw 
(q2) edge[->>, bend left, above] node[above]{$A^1\hspace{.5mm} ,  \hspace{.5mm} A^2$} (q3)
(q3) edge[->>, bend left, above] node[below]{$B^1\hspace{.5mm} ,  \hspace{.5mm} B^2$} (q2); 
\end{tikzpicture}
\end{center}
The  superpotential is 

\begin{eqnarray}
W & \propto & \epsilon_{ij}\epsilon_{kl} \text{Tr}\left[A^iB^kA^jB^l\right].
\end{eqnarray}
The global charges of the conformal field theory are: a $U(1)_R$ factor, two $SU(2)$  factors and finally there is a $U(1)_B$ baryonic symmetry. A fascinating fact about this theory is that it admits a gravity dual in terms of strings in AdS$_5\times T^{1,1}$. The isometries of $T^{1,1}$ realize  the mesonic symmetries of the field theory in terms of the isometries of  $\mathbb{CP}^1 \times \mathbb{CP}^1$; the $U(1)_B$ baryonic symmetry is  associated to the unique non-trivial three-cycle of the geometry. It is worth pointing out that the rotating electrically charged black holes dual to the superconformal index have not yet been constructed on the supergravity side, and that remains an outstanding problem. 

We use the basis for the charges suggested by the toric diagram discussed in \cite{Amariti:2019mgp} and we summarize them in the following table:
\begin{center}
 \begin{tabular}{|c| c| c| c|c|} 
 \hline
 Field & $U(1)_R$ & $U(1)_1$ & $U(1)_2$ & $U(1)_3$ \\ [0.5ex] 
 \hline
 $A_1$ & 1/2 & 1 & 0 & 0 \\ 
 \hline
 $A_2$ & 1/2 & 0 &0& 1 \\
 \hline
 $B_1$ & 1/2 & 0 & 1 & 0 \\
 \hline
$B_2$ & 1/2 & -1 & -1 &- 1 \\ [1ex]
 \hline
\end{tabular}
\end{center}
After performing the shifting $\Delta_{1,2,3} \rightarrow \Delta_{1,2,3} - \frac{\tau}{2}$, we are ready to evaluate equation (\ref{similar}):
\begin{eqnarray}
\log \mathcal{I} & = &- \frac{i  \pi N^2}{3\tau^2}  \left[K\left(\Delta_{1}, \tau\right) + K(\Delta_2, \tau)+ K(\Delta_3, \tau) -K\left(-(- \Delta_1 -\Delta_2 - \Delta_3), \tau\right)\right]\label{similar2} \\ \nonumber
& = & - \frac{i  \pi N^2}{\tau^2}  \left[- \Delta_1^2 \left(\Delta_2 + \Delta_3 \right)- \Delta_2 \Delta_3\left(1 -  2 \tau + \Delta_2 + \Delta_3\right) - \Delta_1 \left(\Delta_2 +\Delta_3 \right)(1- 2\tau + \Delta_2 + \Delta_3)\right].
\end{eqnarray}
 After imposing the condition $\sum_{I=1}^d \Delta_I - 2 \tau = -1$ yields :
\begin{eqnarray}
\log \mathcal{I} & = &  - \frac{i  \pi N^2}{\tau^2}  \left[\Delta_2 \Delta_3 \Delta_4 + \Delta_1 \Delta_3 \Delta_4 +\Delta_1 \Delta_2 \Delta_3+ \Delta_1 \Delta_2 \Delta_4\right]\label{similar21} 
\end{eqnarray}
We see that $\log \mathcal{I}$ presents the behavior proposed in (\ref{similar1}).
\subsection{The Suspended Pinch Point}
The suspended pinch point (SPP) gauge theory corresponds to the near horizon limit of a stack of $N \hspace{2mm} D3$ branes probing the tip of the conical singularity , $x^2 y = w z$. The SPP gauge theory is described by the following quiver 
\begin{center} 
\begin{tikzpicture}
\node[state] at (0,3) (q1) {$N_1$};
\node[state] at (2,0) (q2) {$N_2$};
\node[state] at (-2,0) (q3) {$N_3$};
\draw (q1)  edge[loop above] node{$\phi$} (q1)
(q2) edge[->] node[below]{$X_{32}, X_{23}$} (q3)
(q3) edge[->] node{} (q2)
(q1) edge[->] node[above right]{$X_{12}, X_{21}$} (q2)
(q2) edge[->] node{} (q1)
(q3) edge[->] node[above left]{$X_{31}, X_{13}$} (q1)
(q1) edge[->] node{} (q3);
\end{tikzpicture}
\end{center} 
All the ranks are taken to be the same with $N_1 = N_2 = N_3 =N$ and the sub-indices are meant to help understand the representation properties of the matter fields. The superpotential is 
\begin{eqnarray}
W & = & \text{Tr}\left[X_{21}X_{12}X_{23}X_{32} - X_{32}X_{23}X_{31}X_{13}+X_{13}X_{31}\phi -X_{12}X_{21}\phi\right].
\end{eqnarray}
 Each $X_{ij}$ transforms in the $\mathbf{N}$ representation of the index $i$-th node and in the $\overline{\mathbf{N}}$ of the $j$-th node. The field $\phi$ transforms in the adjoint representation of the corresponding gauge group. The charge assignment for the $U(1)_R$ and the extra $U(1)_i$ global symmetries  can be taken as:
\begin{center}
 \begin{tabular}{|c| c| c| c|c|c|} 
 \hline
 Field & $U(1)_R$ & $U(1)_1$ & $U(1)_2$ & $U(1)_3$ & $U(1)_4$ \\ [0.5ex] 
 \hline
 $\phi$ & 4/5 &1 & 1 &0 &0\\ 
 \hline
 $X_{12}$ & 2/5 & 0 & 0& 0 & 1\\
 \hline
 $X_{21}$ &4/5 & -1 & -1 & 0 & -1 \\
 \hline
$X_{23}$ & 2/5 & 0& 1 & 0 & 0 \\
\hline
$X_{32}$ & 2/5 & 1 & 0 & 0 &0\\
\hline 
$X_{31}$ & 4/5 &- 1 & -1 & -1 &0  \\
\hline
$X_{13}$ & 2/5 & 0 & 0 &1 & 0\\ [1ex]
 \hline
\end{tabular}
\end{center}
We shift now the chemical potentials $\Delta_{1,2,3,4} \rightarrow \Delta_{1,2,3,4} - \frac{2\tau}{5}$. The next step is to use this information and perform the evaluation (\ref{similar}). 
\begin{eqnarray}
\log \mathcal{I} & = &- \frac{i  \pi N^2}{3\tau^2} [K\left(\Delta_{1}+ \Delta_2, \tau\right) + K(\Delta_4, \tau) - K( -(-\Delta_1 - \Delta_2 - \Delta_4), \tau)+K\left(\Delta_2 , \tau\right) +K(\Delta_1,\tau)\label{similarspp} - \\ \nonumber
& &  K( -(-\Delta_1 - \Delta_2 - \Delta_3),\tau) + K(\Delta_4,\tau)]\\ \nonumber
& = & - \frac{i  \pi N^2}{\tau^2}  [- \Delta_1^2\left(\Delta_2 +\Delta_3 + \Delta_4 \right) + \Delta_1 \left((1 - 2 \tau +\Delta_2 +\Delta_3)(-\Delta_2 - \Delta_3) + (2 \tau - 1 - 2 \Delta_2)\Delta_4 - \Delta_4^2\right) + \\ \nonumber
& +& \Delta_2 \left((1 - 2 \tau +\Delta_2)\Delta_3 + \Delta_3^2 + \Delta_4 (1 - 2 \tau + \Delta_2 + \Delta_4)\right) ].
\end{eqnarray}
 Now we use: $\sum_{I =1}^{5} \Delta_I - 2 \tau = -1$ we introduce a fifth fugacity $\Delta_5$ that permits us to rewrite (\ref{similarspp}) in the following, more symmetric,  way:
\begin{eqnarray}
\log \mathcal{I} & = &  - \frac{i  \pi N^2}{\tau^2}  [2 \Delta_2 \Delta_3 \Delta_4 + \Delta_2 \Delta_3 \Delta_5 + \Delta_2 \Delta_4 \Delta_5 + 2 \Delta_1 \Delta_3 \Delta_4 + \Delta_5 \Delta_3 \Delta_2 + \label{spp} \\ \nonumber
& + & \Delta_1 \Delta_4 \Delta_5 + \Delta_1 \Delta_2 \Delta_3 +\Delta_1 \Delta_2 \Delta_4 +\Delta_1 \Delta_2 \Delta_5] 
\end{eqnarray}
This result is in agreement with equation (\ref{resultS}) which is what is expected from toric geometry and reinforces the validity of the analysis  of  \cite{Amariti:2019mgp} which was limited to the Cardy-Like limit.

\subsection{The $\mathbf{dP}_1$ theory}
We consider now the theory arising from a stack of $N \hspace{2mm} D3$ branes at the tip of the complex Calabi-Yau cone whose base is the first del Pezzo surface. The quiver associated to this theory is :
\begin{center} 
\begin{tikzpicture}
\node[state] at (-2,0) (q1) {$N_1$};
\node[state] at (2,0) (q2) {$N_2$};
\node[state] at (2,-4) (q3) {$N_3$};
\node[state] at (-2,-4) (q4) {$N_4$};
\draw
(q2) edge[->>] node[right]{$X^{(\alpha)}_{23}$} (q3)
(q1) edge[->] node[above]{$X_{12}$} (q2)
(q1) edge[->] node[below =.5]{$X_{13}$} (q3)
(q4) edge[->>] node[left]{$X_{41}^{(\alpha)}$} (q1)
(q4) edge[->] node[ above=.5]{$X_{42}$} (q2)
(q3) edge[->>>] node[below]{$X^{(\alpha)}_{34}, X^{(3)}_{34}$} (q4);
\end{tikzpicture}
\end{center}
where $N_1 = N_2 =N_3 =N_4 =N$ and the superpotential is given by:
\begin{eqnarray}
W & = & \epsilon_{\alpha \beta} \text{Tr} \left[X^{(\alpha)}_{34}X^{(\beta)}_{41} X_{13} - X^{(\alpha)}_{34}X^{(\beta)}_{23}X_{42} + X_{12}X^{(3)}_{34}X^{(\alpha)}_{41}X^{(\beta)}_{23}\right].
\end{eqnarray}
The charge assignment specified by the toric data is given by:
\begin{center}
 \begin{tabular}{|c| c| c| c|c|} 
 \hline
 Field & $U(1)_R$ & $U(1)_1$ & $U(1)_2$ & $U(1)_3$ \\ [0.5ex] 
 \hline
 $X_{12}$ & 1/2 &0 & 0 & 1 \\ 
 \hline
 $X^{(1)}_{23}$ & 1/2 & 0 & 1& 0 \\
 \hline
 $X^{(2)}_{23}$ & 1/2 & -1 & - 1 & -1 \\
 \hline
$X^{(1)}_{34}$ & 1 & 0& 1 & 1 \\
\hline
$X^{(2)}_{34}$ & 1 & -1 & -1 & 0 \\
\hline 
$X^{(3)}_{34}$ & 1/2 & 1 & 0 & 0 \\
\hline
$X^{(1)}_{41}$ & 1/2 & 0 & 1 & 0 \\
\hline
$X^{(2)}_{41}$ & 1/2 & -1 &- 1 & - 1 \\
\hline
$X_{13}$ & 1/2 & 1 & 0 & 0 \\
\hline
$X_{41}$ & 1/2 & 1 & 0 & 0 \\ [1ex]
 \hline
\end{tabular}
\end{center}
Let us perform the following transformation of the chemical potentials $\Delta_{1,2,3} \rightarrow \Delta_{1,2,3} - \frac{\tau}{2}$. Evaluating to leading order in $N$ part of the superconformal index according to  (\ref{similar}) :
\begin{eqnarray}
\log \mathcal{I} & = &- \frac{i  \pi N^2}{3\tau^2}  [2 K(\Delta_1,\tau)+K\left(\Delta_{3}, \tau\right) + 2 K(\Delta_2, \tau) - 2K\left(-(- \Delta_1 -\Delta_2 - \Delta_3), \tau\right) + \\ \nonumber
& + & K(\Delta_2+\Delta_3, \tau) - K( -(-\Delta_1 - \Delta_2), \tau)] \label{similar3}  \\ \nonumber
& = &- \frac{i  \pi N^2}{\tau^2}[ -(2 \Delta_1+\Delta_2) \Delta_3^2-3 \Delta_1 \Delta_2 (\Delta_1+\Delta_2+1)- \\ \nonumber
& -&\left(\Delta_2^2+4 \Delta_1 \Delta_2+\Delta_2+2 \Delta_1 (\Delta_1+1)\right) \Delta_3 + \tau (6 \Delta_1 \Delta_2+2 (2 \Delta_1+\Delta_2)
   \Delta_3 ].
\end{eqnarray}
Introducing now $\Delta_4$ via the constraint $\Delta_1 + \Delta_2 + \Delta_3 + \Delta_4 - 2 \tau = -1$ we obtain:
\begin{eqnarray}
\log \mathcal{I} & = &- \frac{i  \pi N^2}{\tau^2}[2 \Delta_1 \Delta_2 \Delta_3 + 3  \Delta_1 \Delta_2 \Delta_4 +2 \Delta_1 \Delta_3 \Delta_4 +  2 \Delta_2 \Delta_3 \Delta_4 ],
\end{eqnarray}
which coincides with the expectation (\ref{resultS}).
\subsection{ $Y^{p,q}$ quiver gauge theories}
The $Y^{pq}$ model corresponds to quiver gauge theories with $2 p$ gauge groups and a chiral field content of bifundamental fields. The charge assignment and the corresponding multiplicity of the fields are shown below:

\begin{center}
 \begin{tabular}{|c| c| c|c|c|} 
 \hline
 Multiplicity  & $U(1)_1$ & $U(1)_2$ & $U(1)_3$ & $U(1)_R$ \\ [0.5ex] 
 \hline
 $p +q$ & 1 & 0 & 0& 1/2 \\ 
 \hline
 $p$ & 0 & 1 &0& 1/2 \\
 \hline
 $p-q$ & 0 & 0 &1 &1/2 \\
 \hline
 $p$ & -1 & -1 &-1& 1/2 \\
\hline
 $ q$& 0& 1& 1& 1 \\
\hline
 $q$ & -1 & -1 &0 & 1 \\  [1ex]
 \hline
\end{tabular}
\end{center}
We proceed to perform the shifting of chemical potentials as follows: $\Delta_{1,2,3} \rightarrow \Delta_{1,2,3} - \frac{\tau}{2}$. Now we evaluate the leading, order ${\cal O}(N^2)$, part of the superconformal index (\ref{similar}):
\begin{eqnarray}
\log \mathcal{I} & = &  -\frac{i\pi  N^2}{3 \tau^2} [(p+q)K \left(\Delta_1, \tau\right) + p K \left(\Delta_2, \tau\right)+(p-q)K \left(\Delta_3, \tau \right)- \label{ypq}\\ \nonumber 
& -  &p K \left(-(- \Delta_1 -\Delta_2 - \Delta_3), \tau\right)+ q K \left(\Delta_{2} + \Delta_3,\tau\right)  -q K \left(-(-\Delta_{1}-\Delta_2),\tau   \right)  ] \\ \nonumber
& = & -\frac{i\pi  N^2}{3 \tau^2}[\Delta_2 (\Delta_1 -\Delta_3) (-q) (\Delta_1 + \Delta_2 + \Delta_3 - 2\tau +1)- \\ \nonumber 
& - & p
   \left((\Delta_2 +\Delta_3 \Delta_1^2+(\Delta_2 + \Delta_3 \Delta_1 (\Delta_
  2+ \Delta_3 -2 \tau +1 )+ \Delta_2  \Delta_3 (\Delta_2 + \Delta_3 - 2 \tau +1)\right)]
\end{eqnarray}
 Finally we eliminate $\tau$ from (\ref{ypq}) using $\sum_{I =1}^{4}\Delta_I - 2 \tau = -1$, which successfully reproduce the structure of (\ref{similar1}):
\begin{eqnarray}
\label{Eq:Ypq}
\log \mathcal{I} & = &  -\frac{i\pi  N^2}{3 \tau^2}[p \Delta_1 \Delta_2 \Delta_3 +(p +  q) \Delta_1 \Delta_2 \Delta_4 +  p \Delta_1 \Delta_3 \Delta_4 + (p - q)\Delta_2 \Delta_3 \Delta_4 ] .
\end{eqnarray}


\section{Corrections to the dual black hole entropy of the conifold theory} \label{Sec:Entropy}
Thus far we have been focused on a specific region of chemical potentials that permited us to simplify all the computations associated with the function $\left[\Delta\right]_{\tau}$. Ultimately, it is necessary to identify the quantity $\log \mathcal{I}$ with the corresponding entropy function for the dual black hole gravity solution. The prototypical example is provided by $\mathcal{N} =4$ SYM, as discussed in \cite{Benini:2018ywd}, where the identification states:
\begin{eqnarray}
\log \mathcal{I} & = & - \frac{i  \pi N^2}{6\tau^2} C_{IJK} \Delta_I \Delta_J \Delta_K \Longleftrightarrow S_{E} = - \frac{i  \pi N^2}{6\tau^2} C_{IJK} X_I X_J X_K,
\end{eqnarray}
provided $\Delta_I \Longleftrightarrow X_I$. More generically, $\left[\Delta_I\right]_{\tau} \Longleftrightarrow X_I$ where $\tau$ is the chemical potential associated to the two equal angular momenta $J_1 = J_2 =J$ and $X_I$ are the chemical potentials associated to the cherges $Q_I$.
As discussed in \cite{Benini:2018ywd}, the identification above is valid provided $X_I$ is within the same analyticity domain that $\left[\Delta\right]_{\tau}$. This is  the case, since both belong to the domain specified by (\ref{const}),  and they satisfy the constraint : 
\begin{eqnarray}
\sum_{I=1}^d X_I -2 \tau & = & -1.
\end{eqnarray}
\\ \par

 The next step would be to obtain the black hole entropy by extremizing the Legendre transform of $\log \mathcal{I}$ with the appropriate constraint:
\begin{eqnarray}
S \left(Q ,  \Lambda\right) & = & S_E + 2  \pi  i \left(\sum_{I = 1}^{d } X_{I} Q_{I} -  2 \tau J\right) + 2 \pi i \Lambda \left(\sum_{I= 1}^{d }X_{I} - 2  \tau + 1\right) ,
\end{eqnarray}
where $\Lambda$ is a Lagrange multiplier imposing the constraint. The black hole entropy can be thus obtained exactly as done in \cite{Cabo-Bizet:2018ehj, Amariti:2019mgp}. In particular the result obtained in \cite{Lanir:2019abx}, which appeared after the first version of our manuscript , confirms our results. Going away from the region of chemical potentials we have been restricting ourselves to so far would produce some modifications in the result.The technical reason being that, in different regions of the domain of complex chemical potentials, the functions $\left[\sum_I q_I \Delta_I\right]_{\tau} $ and $\sum_I q_I\left[\Delta_I\right]_{\tau} $ do not agree.  \\ \par 

The structure (\ref{resultS}) can be lost either by spoiling the homogeneity of $\log \mathcal{I}$ with respect to $\Delta_I$ and $\tau$ or by failing to completely cancel the term that only depends on $\tau$ and comes from the contribution of vector multiplets. Homogeneity in $\Delta_I$ and $\tau$ is crucial to perform the extremization procedure, whereas the appearance of an extra term exclusively dependent on $\tau$ could be easily incorporated in order to explore possible modifications to black the hole entropy. In the particular case of the conifold theory that we studied in section (\ref{conifold}) we have that, for chemical potentials in the region:
\begin{eqnarray}
\text{Im}\left(-\frac{2}{\tau}\right) & > &\text{Im}\left(\frac{\left[\Delta_1\right]_{\tau}+\left[\Delta_2\right]_{\tau}+\left[\Delta_3\right]_{\tau}}{\tau}\right)>\text{Im}\left(-\frac{1}{\tau}\right)\label{const2} \\ \nonumber
& \Downarrow & \\ \nonumber
\left[\Delta_1 +\Delta_2+\Delta_3 \right]_{\tau} & = & \left[\Delta_1\right]_{\tau}+\left[\Delta_2\right]_{\tau}+\left[\Delta_3\right]_{\tau} + 1.
\end{eqnarray}
therefore, the superconformal index takes the following form:
\begin{eqnarray}
\log \mathcal{I} & = &  - \frac{i  \pi N^2}{\tau^2}  \{\left(\left[\Delta_2\right]_{\tau} + \frac{1}{2}\right)\left(\left[\Delta_3\right]_{\tau} + \frac{1}{2}\right)\left(\left[\Delta_4\right]_{\tau} + \frac{1}{2}\right)+ \label{similar3}  \\ \nonumber
& + & \left(\left[\Delta_1\right]_{\tau} + \frac{1}{2}\right)\left(\left[\Delta_3\right]_{\tau} + \frac{1}{2}\right)\left(\left[\Delta_4\right]_{\tau} + \frac{1}{2}\right)+\\ \nonumber
& + &\left(\left[\Delta_1\right]_{\tau} + \frac{1}{2}\right)\left(\left[\Delta_2\right]_{\tau} + \frac{1}{2}\right)\left(\left[\Delta_3\right]_{\tau} + \frac{1}{2}\right)+ \\ \nonumber
& + & \left(\left[\Delta_1\right]_{\tau} + \frac{1}{2}\right)\left(\left[\Delta_2\right]_{\tau} + \frac{1}{2}\right)\left(\left[\Delta_4\right]_{\tau} + \frac{1}{2}\right)\} - i \pi N^2 \left(\frac{1}{2 \tau} -1\right)
\end{eqnarray}
\\ \par 
Notice that the appearance of a contribution that only depends on $\tau$ in this case is related to the specific details of how the function $[\cdots]_{\tau}$ behaves in the different domains of chemical potentials. The case of $\mathcal{N}=4$ SYM is special because the  function $\log \mathcal{I} \sim \Delta_1 \Delta_2 \Delta_3$ is quite simple and consists only of one term.  For this reason one could hope to eliminate all contributions depending only on $\tau$ by modifying the constraint obeyed by the chemical potentials $\sum_{I=1}^d \Delta_I - 2 \tau = -1 \rightarrow \sum_{I=1}^d \Delta_I - 2 \tau = 1$. This is, indeed,  the case verified in \cite{Benini:2018ywd}.  However, for more complicated theories where $\log \mathcal{I}$ has a more than one term dictated by the anomaly coefficients $C_{IJK}$,  the extra pice persists as we see in \eqref{similar3}.

\begin{figure}[H]
\caption{The figure shows the complex plane of chemical potentials for generic $\Delta$ including also the region for the corresponding $X_I$ inside the dashed strip. notice that the grey and the dashed region overlap in a zone where the identification $\left(\left[\Delta_I\right]_{\tau} + \frac{1}{2}\right) \Longleftrightarrow X_I $ is valid. }
\centering
    \includegraphics[width=1\textwidth]{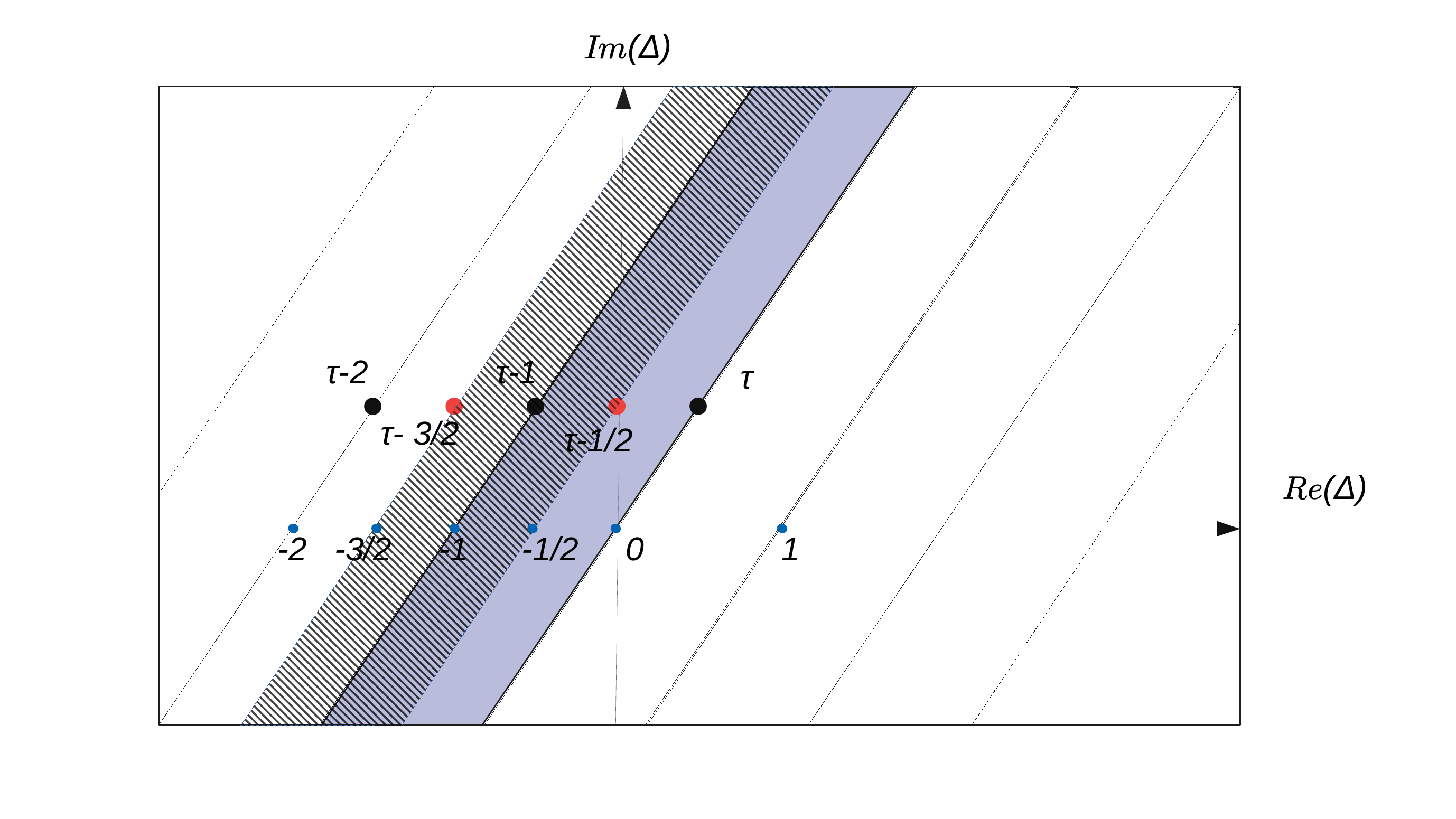}
    \label{region}
    \end{figure}

Let us now investigate how this extra term  modifies the entropy obtained by taking the Legendre transform of $\log \mathcal{I}$. Our starting point it to organize the computation as to maximally take advantage of the scaling properties   of $S_E$, which implies that now we propose the identification $\left(\left[\Delta_I\right]_{\tau} + \frac{1}{2}\right) \Longleftrightarrow X_I $ within the region (\ref{const2}). Notice that now the constraint is modified as:
\begin{eqnarray}
\sum_{I=1}^4\left[\Delta_I\right]_{\tau} - 2 \tau & = & -1 \label{modif} \\ \nonumber
& \Downarrow & \\ \nonumber 
\sum_{I=1}^4\left(X_I -\frac{1}{2} \right)- 2 \tau & = & -1  \\ \nonumber
& \Downarrow & \\ \nonumber
\sum_{I=1}^4X_I - 2 \tau & = & 1 .
\end{eqnarray}
Even though the constraint (\ref{modif}) has been modified when identifying $\left(\left[\Delta_I\right]_{\tau} + \frac{1}{2}\right) \Longleftrightarrow X_I $ we notice that the new constraint corresponds precisely to the other possible choice of relation among the chemical potentials as discussed in \cite{Cabo-Bizet:2019osg, ArabiArdehali:2019tdm}.
If the chemical potentials are in the region (\ref{const2}), then :
\begin{eqnarray}
\text{Im}\left(-\frac{2}{\tau}\right) & > &\text{Im}\left(\frac{X_1 +X_2 + X_3}{\tau} -\frac{3}{2 \tau}\right)>\text{Im}\left(-\frac{1}{\tau}\right) \label{regionX} \\ \nonumber
& \Downarrow & \\ \nonumber
\text{Im}\left(-\frac{1}{2 \tau}\right) & > &\text{Im}\left(\frac{X_1 +X_2 + X_3}{\tau} \right)>\text{Im}\left(\frac{1}{2\tau}\right).
\end{eqnarray}
Note that this region does not coincides precisely with the fundamental domain over which the $X_I$ are defined (\ref{const}), however there is a non-empty intersection between the two as we illustrate in Fig.\ref{region}. 

Thus we write:
\begin{eqnarray}
\widehat{S} & = & S_E +S_{\tau} \\ \nonumber
S_E& = & -\frac{i \pi N^2}{ \tau^2}\left(X_1 X_2 X_3 + X_1 X_2 X_4 + X_1 X_3 X_4 + X_2 X_3 X_4 \right) \label{anabel} \\ \nonumber
S_{\tau} & = &  \frac{i \pi N^2}{\tau^2} \tau \left(\tau-\frac{1}{2 }\right) 
\end{eqnarray}

Since $S_\tau$ is independent of   $X_I$ we have:
\bea
 \frac{\partial \widehat{S}}{\partial  X_I}  & = & \frac{\partial  S_E}{\partial X_I} \\ \nonumber
 \frac{\partial  \widehat{S}}{\partial \tau}  & = & \frac{\partial  S_E}{\partial \tau} + \frac{\partial S_\tau}{\partial \tau} \\ \nonumber
\end{eqnarray}
The function we need to extremize now is the following:
\begin{eqnarray}
S \left(Q ,  \Lambda\right) & = & \widehat{S}+ 2  \pi  i \left(\sum_{I = 1}^{d } X_I Q_{I} -  2 \tau J\right) + 2 \pi i \Lambda \left(\sum_{I= 1}^{d }X_I - 2  \tau - 1\right) ,
\end{eqnarray}

 The extremization condition implies:
\begin{eqnarray}
\frac{\partial S}{\partial  X_I} & = & 0 , \hspace{2mm} \Rightarrow \hspace{2mm} \frac{\partial   \widehat{S}}{\partial  X_I} = - 2  \pi i \left(Q_I + \Lambda\right) \\ \nonumber
\frac{\partial S}{\partial \tau} & = & 0  \hspace{2mm} \Rightarrow \hspace{2mm} \frac{\partial S_E}{\partial \tau} =  - 4  \pi i \left(\widetilde{J}-\Lambda \right) \\ \nonumber
\widetilde{J} & \equiv &  J + \frac{1}{4 \pi i } \frac{\partial S_{\tau}}{\partial \tau} .
\end{eqnarray}
The homogeneity of $S_E$ leads to the important relation:
\begin{eqnarray}
S_E & = & \sum_{I=1}^{d}  X_I \frac{\partial S_E}{\partial X_I} +  \tau \frac{\partial S_E}{\partial \tau }. \label{see}
\end{eqnarray}
Following \cite{Amariti:2019mgp}, we insert (\ref{see}) in  (\ref{anabel}) and evaluating on the extremization solutions we find:
\begin{eqnarray}\label{Eq:EntropCor}
S \left(Q, J\right) & = &  2  \pi i \Lambda(Q,J)  + S_{\tau} -4\pi i  \tau \left(\widetilde{J} - J\right) \\ \nonumber
& = & 2 \pi i \Lambda(Q,J)  + S_{\tau} -  \tau \frac{\partial S_{\tau}}{\partial \tau} \\ \nonumber
& = & 2 \pi i \Lambda(Q,J) +  \frac{ i  \pi N^2}{\tau(J)}\left(\tau(J) -1 \right) .
\end{eqnarray}
 For a particular set of values of $X_I $,  the properties of $S_E$ allow us to reconstruct $(S_E)^2$   from suitable combinations of products of its derivatives with respect to $X_I $ which generically leads to a cubic equation to determine $\L(Q,J)$.  If we choose 
$ X_1 = X_3  $  , then $S_E$ for the conifold theory coincides with $S_E$ for the $Y^{p,p}$ theory described in \cite{Amariti:2019mgp}, namely:
\begin{eqnarray}
 \widetilde{S}_E \equiv S_E\big{|}_{ X_1 = X_3} & = & -\frac{i \pi N^2}{ \tau^2}\left(X_1^2X_2+2X_1X_2X_4+X_1^2X_4 \right)\label{Yppconifold}.
\end{eqnarray}
Now we can follow the extremization procedure put forward in \cite{Amariti:2019mgp} keeping track of the correction $S_{\tau}$ when taking $Q_3 \rightarrow Q_1$ for the $Y^{1,1}$ quiver gauge theory. It can be shown that $\widetilde{S}_E$ satisfies:
\begin{eqnarray}
0 & = & \frac{\partial \widetilde{S}_E}{\partial X_1}\left[2 \left(2 \frac{\partial \widetilde{S}_E}{\partial  X_1} + \frac{\partial \widetilde{S}_E}{\partial X_4}\right)\frac{\partial \widetilde{S}_E}{\partial X_2 } - \left(\frac{\partial \widetilde{S}_E}{\partial X_2}\right)^2 - \left(2 \frac{\partial \widetilde{S}_E}{\partial X_1} - \frac{\partial \widetilde{S}_E}{\partial X_4}\right)^2\right]  \\ \nonumber 
& + & 4 p N^2 \left(\frac{\partial \widetilde{S}_E}{\partial \tau}\right)^2.
\end{eqnarray}
Using equation (\ref{anabel}), we can obtain a cubic equation for $\Lambda$ in the same spirit as \cite{Amariti:2019mgp} and also to keep track of the modification produced in the entropy by the presence of  $S_{\tau}$ in  (\ref{similar1}), hence, we have:
\begin{eqnarray}\label{Lambda}
0 & = & (Q_1 + \Lambda) [2 \left(2 (Q_1 + \Lambda) + (\Lambda + Q_4)\right)(\Lambda + Q_2) - (\Lambda + Q_2)^2 - \\ \nonumber
& - & \left(2(\Lambda + Q_1) - (\Lambda + Q_4)\right)^2] + 4 p N^2 (\Lambda - \widetilde{J})^2. 
\end{eqnarray}
Equation (\ref{Lambda}) can be written as:
\begin{eqnarray}
0 & = & \Lambda^3 + \widetilde{p}_2 \Lambda^2 +\widetilde{ p}_1 \Lambda + \widetilde{p}_0  \label{cubic} \\ \nonumber
\widetilde{p}_0 & = &  N^2 p \widetilde{J}^2- \frac{1}{4}Q_1 Q_2^2 + Q_1^2Q_2+\frac{1}{2}Q_1 Q_2 Q_4 -Q_1^3-\frac{1}{4}Q_1 Q_4^2+Q_1^2Q_4\\ \nonumber
\widetilde{p_0}|_{\widetilde{J}= J} & \equiv & p_0 \\ \nonumber
\widetilde{p}_1 & \equiv & p_1 - \frac{N^4}{4 \tau^2}= 2 Q_1 (Q_2 + Q_4) + \frac{Q_4 Q_2}{2} - \frac{Q_2^2}{4} - Q_1^2 - \frac{Q_4^2}{4} - 2 N^2  \widetilde{J}\\ \nonumber
\widetilde{p}_2 & \equiv & p_2 =  N^2 p +2 Q_1 +Q_2.
\end{eqnarray}
Demanding the condition
\begin{eqnarray}
 \widetilde{p}_0 = \widetilde{p}_1 \widetilde{p}_2, \label{condp0p1p2}
\end{eqnarray} 
 the assumption of real charges in \cite{Cabo-Bizet:2018ehj, Amariti:2019mgp} led to purely imaginary values of $\Lambda$ and therefore to a real entropy. We need to be more careful since \eqref{Lambda} is a modified version of the one appearing in \cite{Amariti:2019mgp}. The modifications enter through $\widetilde{J}$ and $\frac{ i  \pi N^2}{\tau(J)}\left(\tau(J) -1 \right)$ in \eqref{Eq:EntropCor}. Let us still demand the condition \eqref{condp0p1p2}, which a priory do not ensure real entropy but gives a simplified enough expression that we can work with. Reality of the entropy then would impose that, separately, the correction was a real number:
 \begin{eqnarray}
  \text{Re}\left( \frac{ 1}{\tau(J)}\left(\tau(J) -1 \right)\right) & = & 0. \label{reality}
 \end{eqnarray}
Imposing \eqref{reality} would constraint the set of possible values $\tau$ could take. Eve though in some contexts \cite{Benini:2018ywd, Lanir:2019abx} the values of $\tau$ for which one can obtain a reasonable black hole entropy are constrained, we do not have any a priory reason for which $\tau$ should satisfy \eqref{reality}. Of course, a more rigorous approach is required, since the reality condition for the full entropy would imply a relation among the coefficients \eqref{cubic} far more complicated than \eqref{condp0p1p2}. However, at least for the values of $\tau$ that ensure reality of \eqref{Eq:EntropCor}, through \eqref{condp0p1p2} and \eqref{reality} we can proceed as follows.

The solution of (\ref{Lambda}) when plugged into equation (\ref{Eq:EntropCor}) leads to an entropy of the form: 
\begin{eqnarray}
S \left(Q ,\Lambda\right) & = & 2 \pi \sqrt{2 Q_1 (Q_2 + Q_4) + \frac{Q_4 Q_2}{2} - \frac{Q_2^2}{4} - Q_1^2 - \frac{Q_4^2}{4} - 2 N^2  \widetilde{J}} \\ \nonumber
& + & \frac{ i  \pi N^2}{\tau(J)}\left(\tau(J) -1 \right) \\ \nonumber
& = & 2 \pi \sqrt{2Q_1(Q_2 + Q_4) + \frac{Q_4 Q_2}{2} - \frac{Q_2^2}{4} - Q_1^2 - \frac{Q_4^2}{4} - 2 N^2 \left[J +\frac{N^2}{8\tau(J)^2}\right] } \\ \nonumber
& + & \frac{ i  \pi N^2}{\tau(J)}\left(\tau(J) -1 \right)
\end{eqnarray}
In the above expression we have left $\widetilde{J} =J+\frac{1}{4 \pi i}\frac{\partial S_{\tau}}{\partial \tau}$ explicitly in the corrections to highlight its effect.  The angular velocity $\tau(J)$ appears only formally, it should be substituted by the extremization procedure, we have indicated such operation as $\tau(J)$.  The most dramatic effect is a shift in the angular momentum. 

\section{Conclusions}\label{Sec:Conclusions}

In this brief note we have  explored the superconformal index following the Bethe Ansatz approach introduced by Benini and Milan \cite{Benini:2018mlo}. We have shown that a class of solutions can be extended to solve the Bethe Ansatz equation for a large class of 4d ${\cal N}=1$ supersymmetric gauge  theories. The Bethe Ansatz approach has the advantage that it does not require to take the Cardy limit and therefore provides a more complete large $N$ expression.  Indeed, for generic toric quiver gauge theories we determined that  there is a region in the space of chemical potentials in which the  ${\cal O}(N^2)$  result obtained in the cardy-like limit can be recovered buttessing previous results in the literature \cite{Cabo-Bizet:2019osg, Amariti:2019mgp, Kim:2019yrz}. Furhtermore, at least for the simple case of the conifold theory we saw that one can obtain a similar structure of the superconformal index with extra corrections in $\tau$ , but sufficiently simple as to permit us to proceed with the extremization procedure and consequently a corrected black hole entropy.   We hope that more work along this direction might eventually allow to understand the growth of states in the index in a more systematic fashion that covers all the possible regions in the space of chemical potentials. For example, by exploiting the Bethe Ansatz approach to the topologically twisted index a systematic study of $1/N$ corrections for the ABJM index was performed in \cite{Liu:2017vll}; a similar study for a Chern-Simons matter theory dual to massive IIA black holes was  reported in \cite{Liu:2018bac}. Such understanding of $1/N$ corrections will naturally translate into interesting aspects in the dual quantum gravity side for AdS$_5$ black holes.  For example, the statistical entropy of certain magnetically charged  AdS$_4$ black holes has recently been given a microscopic explanation in terms of the topologically twisted index \cite{Benini:2015eyy} (see \cite{Hosseini:2018qsx,Zaffaroni:2019dhb} for a reviews with comprehensive lists of references).  The investigation of sub-leading (logarithmic in $N$) corrections such as those performed recently \cite{Liu:2017vbl,Gang:2019uay} have helped clarify the nature of the degrees of freedom on the gravitational side of the duality.  One would hope for  similar developments  in the context of AdS$_5$ black holes. 

There are many other  interesting open problems. At the technical level, it would be interesting to generalize the Bethe Ansatz approach to arbitrary fugacities such that a general expression depending on both angular momenta can be achieved. There is little doubt that such generalization will yield the expected results but it will clarify the inner workings of the evaluation of the superconformal index.  In this manuscript we have completely avoided the subtle discussion concerning the space of solutions of the Bethe Ansatz equations, we limited ourselves to just one class and showed that it yields a contribution sufficient to extract the dual black hole entropy and its potential corrections in the appropriate domain of chemical potentials. It would be very illuminating to have a better understanding of all the solutions and how one should weight their contributions to the index. 

Finally, it is an important open problem to construct explicitly the black holes dual to the field theories discussed in this manuscript. Our computation, as well as those in a number of recent publications \cite{Cabo-Bizet:2019osg,Kim:2019yrz,Amariti:2019mgp}, show that it is relatively easy to find the superconformal index in a large class of supersymmetric four-dimensional field theories some of which have known supergravity dual. Moreover, using the entropy formula one can evaluate the entropy and realize that it corresponds to that of large black holes  in  AdS$_5$.  However, the explicit black hole construction on the gravity side is still in its infancy, not much is known beyond the AdS$_5$ black holes dual to ${\cal N}=4$  SYM (and some of its orbifolds). It remains an outstanding challenge for the supergravity community to explicitly construct rotating electrically charged black holes which could be understood as dual of available field theory results. One particular example that comes to mind among the class discussed in this note  would be the black holes in asymptotically AdS$_5\times T^{1,1}$ and, more generally, AdS$_5\times Y^{p,q}$.

\section*{Acknowledgments}
We are thankful to  Antonio Amariti, Francesco Benini, Alejandro Cabo-Bizet,  Ivan Garozzo, Gabriele Lo Monaco, Jun Nian, Paolo Milan  and Alberto Zaffaroni. We thank the anonymous JHEP referee who suggested important improvements to the original version.  LPZ is partially supported by the U.S. Department of Energy under grant DE-SC0007859.

\bibliographystyle{JHEP}
\bibliography{BHLocalization}
\end{document}